\documentclass{elsart}

\usepackage{natbib}

\usepackage{epsfig}

\usepackage{amssymb}

\def\Chandra{{\it Chandra }}
\def\XMM{{\it XMM }}

\def\WMAP{{\it WMAP }}

\def\mnras{MNRAS}

\def\apj{ApJ}
\def\apjs{ApJS}
\def\aap{A\&A}

\def\newa{NewA}

\def\h2gcm3{h^{2} \, \hbox{g}\,\hbox{cm}^{-3}}
\def\Lxunit{\, \hbox{h}^{-2} \, \hbox{erg} \, \hbox{s}^{-1}}
\def\kevcm2{\, \hbox{keV} \, \hbox{cm}^{2}}

\def\hMpc{\,h^{-1}\,\hbox{Mpc}}

\def\hMsol{h^{-1}\,\hbox{M}_{\odot}}

\def\ergscm2{\,\hbox{erg}\,\hbox{s}^{-1}\hbox{cm}^{-2}}

\def\keV{\,\hbox{keV}}
\def\icm2{\,\hbox{cm}^{-2}}
\def\icm3{\,\hbox{cm}^{-3}}
\def\gcm3{\,\hbox{g}\,\hbox{cm}^{-3}}

\def\hkpc{\,h^{-1}\,\hbox{kpc}}

\def\h50{\, h_{50}}

\def\sl0{\epsilon_{0}}

\def\H0{H_0=100 \, h \, {\rm kms^{-1}Mpc^{-1}}}
\def\Zsol{Z_{\odot}}

\def\log{{\rm log}}

\def\AP3M{$\rm{AP}^3\rm{M}$\,}
\def\p3m{{\rm P}$^{3}${\rm M}\,}
\def\ap3m{{\rm AP}$^{3}${\rm M}\,}

\def\rhocr0{\rho_{\rm cr,0}}

\def\etal{{et al.\thinspace}}
\def\eg{{e.g.\thinspace}}

\def\LCDM{\Lambda \hbox{CDM}}
\def\CLEF{{CLEF }}

\begin{document}

\begin{frontmatter}



\title{Clusters of Galaxies: New Results from the \CLEF Hydrodynamics Simulation}
\author[Sussex]{S.T.~Kay\corauthref{cor}},
\corauth[cor]{Corresponding author.}
\ead{s.t.kay@sussex.ac.uk}
\author[IAS]{A.C.~da Silva},
\author[IAS]{N.~Aghanim},
\author[Toulouse]{A.~Blanchard},
\author[Sussex]{A.R.~Liddle},
\author[IAS]{J.-L.~Puget},
\author[Toulouse]{R.~Sadat} and
\author[Sussex]{P.A.~Thomas}
\address[Sussex]{Astronomy Centre,
                 University of Sussex,
                 Falmer, Brighton BN1 9QH, UK}
\address[IAS]   {IAS, B\^atiment 121,
                 Universit\'e Paris Sud, 
                 F-91405 Orsay, France}
\address[Toulouse]{Observatoire Midi-Pyr\'en\'ees, 
                   Av. Edouard Belin 14, 
                   F-31500 Toulouse, France}
\begin{abstract}
Preliminary results are presented from the CLEF 
hydrodynamics simulation, a large ($N=2\times 428^3$
particles within a $200\hMpc$ comoving box) simulation 
of the $\LCDM$ cosmology that includes both radiative 
cooling and a simple model for galactic feedback. 
Specifically, we focus on the X-ray properties of the
simulated clusters at $z=0$ and demonstrate a reasonable
level of agreement between simulated and observed cluster
scaling relations.
\end{abstract}


\end{frontmatter}

\section{Introduction}
\label{sec:intro}
As the largest and latest virialised structures to form,
galaxy clusters are especially useful cosmological probes
\citep[\eg see][and references therein]{Viana03}.
Next generation cluster cosmology surveys, such as the XCS 
\citep{Romer01}, will detect sufficiently large numbers 
of clusters that uncertainties in values of cosmological parameters 
will be mainly systematic, requiring for example an
accurate calibration between cluster X-ray temperature and
mass. Such measurements demand an improved understanding of 
cluster physics, therefore realistic numerical simulations of
the cluster population are essential.

In this paper we present a preliminary analysis of the 
$z=0$ cluster population within the CLEF hydrodynamics 
simulation, a large state-of-the-art cosmological
simulation that, besides gravity and gas dynamics, includes a model 
for the effects of galaxy formation. As we will show, the simulation 
does a reasonably good job at reproducing X-ray scaling relations at 
$z=0$.

\section{The CLEF hydrodynamics simulation}
\label{sec:sim}

\begin{figure}
\centering
\centerline{\epsfig{file=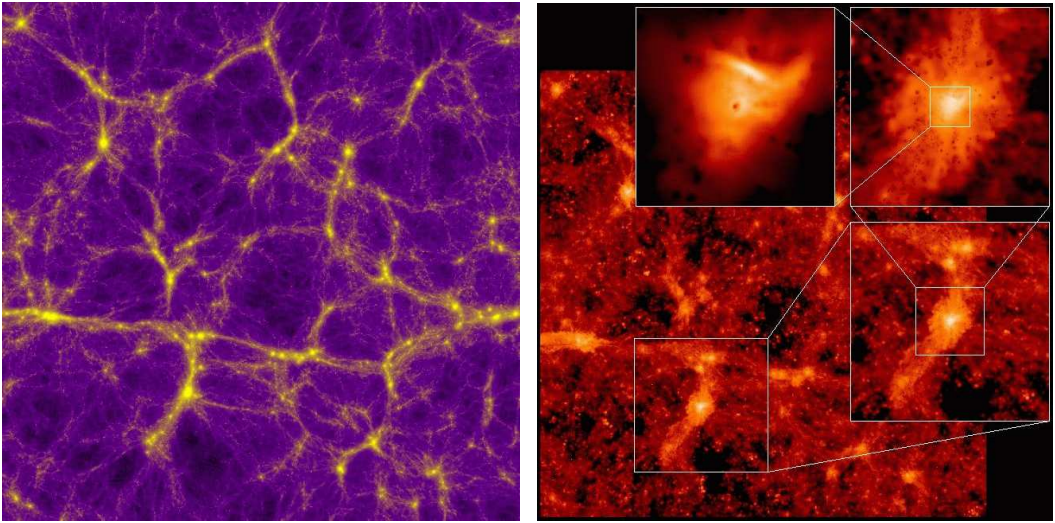,width=5.5in}}
\caption{Left: optical depth image of the gas in a 
$200\times200\times10\hMpc$ slice at $z=0$. Right: series
of zooms showing images of the mass-weighted temperature 
of the gas, from the full box width to an individual
cluster.}
\label{fig:image}
\end{figure}

The CLEF (CLuster Evolution and Formation) hydrodynamics
simulation (see Fig~\ref{fig:image}) is a large simulation of structure 
formation within the $\Lambda$CDM cosmology, with the following
cosmological parameters: $\Omega_{\rm m}=0.3$, $\Omega_{\Lambda}=0.7$, 
$\Omega_{\rm b}h^{2}=0.0238$, $h=0.7$ and $\sigma_8=0.9$. 
These values are in good agreement with recent \WMAP 
analyses \citep{Spergel03}. 

Initial conditions were generated using a modified version of the {\sc
cosmic} software package provided with the {\sc hydra} code
\citep{Couchman95}.  The appropriate transfer function, generated
using {\sc cmbfast} \citep{Seljak96}, was read in and a displacement
field generated for a $200\hMpc$ comoving box at $z=49$. Two regular
cubic grids of $428^3$ particles, separated by half the interparticle
distance in each of the $x, y$ and $z$ directions, were then perturbed
by these displacements to create the initial particle positions.
Thus, the gas and dark matter particle masses were set to $m_{\rm
gas}=1.4\times 10^{9}\hMsol$ and $m_{\rm dark}=7.1\times 10^{9}\hMsol$
respectively.

This initial configuration was then evolved to $z=0$ using version
2 of the {\sc gadget} code \citep{Springel01}, a hybrid Particle-Mesh/Tree
gravity solver with a version of Smoothed Particle Hydrodynamics (SPH)
that explicitly conserves entropy where appropriate. 
In addition, the gas could cool radiatively, 
assuming a fixed metallicity of $Z=0.3\Zsol$. Cooled gas, with 
$n_{\rm H}>10^{-3}\icm3$ and $T<1.2\times 10^{4}$K, could either form
stars if $r>f_{\rm heat}$ or be reheated by stars if $r<f_{\rm heat}$,
where $r$ is a random number drawn for each particle from the unit interval 
and $f_{\rm heat}=0.1$
is the reheated mass fraction parameter. Each reheated gas particle 
was given a fixed amount of entropy, $S_{\rm heat}=1000 \kevcm2$,
where $S \equiv kT/n^{2/3}$, which further heats the ICM
as the particle does work on its surroundings.
Further details may be found in \citet{Kay04}.

\section{X-ray scaling relations at $z=0$}
\label{sec:srel}

In this paper, we concentrate on comparing a selection of simulated
and observed X-ray cluster scaling relations at $z=0$. Clusters were
identified by first identifying local maxima in the density field
and growing spheres around these maxima until the average density
within each sphere was a fixed factor, $\Delta$, above the critical
density, $\rho_{\rm cr}=3H_{0}^2/8\pi G$. Values of $\Delta$ used
will be given in each subsection.
For the virial density ($\Delta\sim 104$)
there are $>400$ clusters with $kT_{\rm vir}>1$ keV 
($>60$ above 3 keV).

\subsection{Temperature--mass relation}
\label{subsec:tm}

\begin{figure}
\centering
\centerline{\epsfig{file=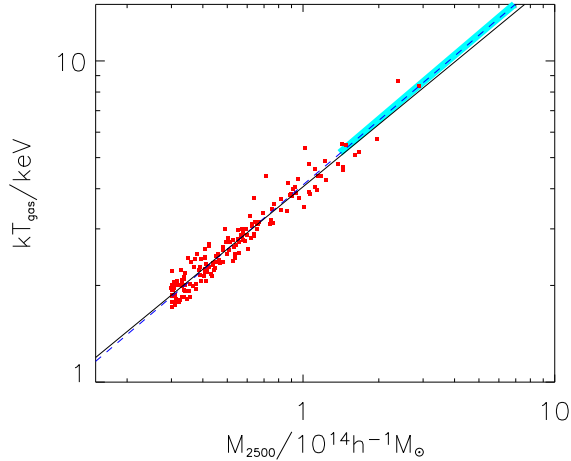,width=3in}}
\caption{Gas mass-weighted temperature versus mass,
evaluated at $\Delta=2500$. The dashed line is
the best-fit relation with the self-similar slope 
$2/3$. The solid line is the best-fit relation,
allowing both the normalisation and slope to vary. The
solid band is the best-fit relation to clusters studied
by Allen, Schmidt and Fabian (2001).}
\label{fig:tm2500}
\end{figure}

\vspace{-0.5cm}
We begin by showing in Fig.~\ref{fig:tm2500} the relation 
between hot gas mass-weighted temperature 
($T_{\rm gas} \equiv \Sigma_i m_i T_i / \Sigma_i m_i$, where
the sum is over all gas particles with $T_i>10^{5}$K) and
total mass for a density contrast $\Delta=2500$. All clusters
with $M_{2500}>3\times 10^{14}\hMsol$ are considered. The
dashed line is a best-fit relation to the clusters for 
a fixed slope of 2/3, as expected if the clusters form a 
self-similar population. This relation is
\begin{equation}
\log(kT_{\rm gas}/\keV) = (0.614 \pm 0.003) + 
(2/3) \, \log(M_{2500}/M_{14}),
\label{eqn:tm_ss}
\end{equation}
where $M_{14}=10^{14}\hMsol$.
When the slope is allowed to vary, the best-fit relation
(solid line) is
\begin{equation}
\log(kT_{\rm gas}/\keV) = (0.608 \pm 0.004) + 
(0.65 \pm 0.01) \, \log(M_{2500}/M_{14}),
\label{eqn:tm_bf}
\end{equation}
close to the self-similar relation. The thick band is the observed 
relation derived by \citep{Allen01}, in good agreement with our results.

It is more common in the literature for observed temperature-mass
relations to be presented at larger radii, using spectroscopic
(photon-weighted) temperatures and mass estimates assuming
$\beta$-model surface brightness profiles and polytropic-model
temperature profiles (e.g. \citealt{Nevalainen00};
\citealt{Finoguenov01}; \citealt{Sanderson03}). These results
generally suggest a slope closer to 1/2 than 2/3, attributed to
non-gravitational processes (see below), and a normalisation that is
offset in mass from simulation predictions by $\sim$40 per cent.
Examining, for example, the X-ray emission-weighted temperature-mass
relation from our simulation at $r_{500}$, $T_{\rm X}-M_{500}$
[$T_{\rm X} \equiv \Sigma_i m_i n_i \Lambda(T_i) T_i / \Sigma_i m_i
n_i \Lambda(T_i)$, where $\Lambda$ is an energy-dependent cooling
function], we find a similar slope to the observations (0.53) but an
offset in normalisation comparable to previous simulations. The cause
of this offset is likely due to incorrect estimates of cluster masses
\citep[e.g.][]{Rasia04} and is something we will return to in a future
paper.

\subsection{Entropy--temperature relation}
\label{subsec:st}

\begin{figure}
\centering
\centerline{\epsfig{file=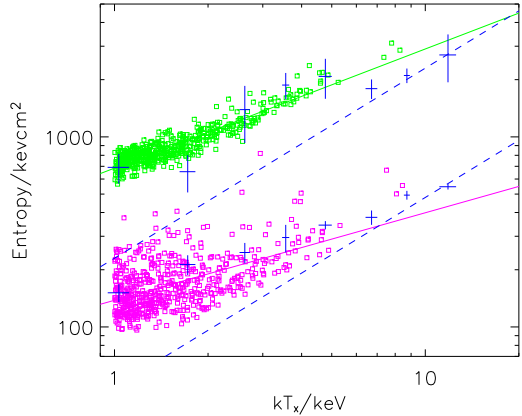,width=3in}}
\caption{Entropy versus X-ray emission-weighted temperature at
$0.1r_{200}$ (lower points) and $r_{500}$ (upper points), with the 
solid lines being fits to these data. Crosses
are data from Ponman, Sanderson and Finoguenov (2003) and dashed lines
are self-similar scalings, normalised to their hottest clusters.}
\label{fig:st}
\end{figure}

\vspace{-0.5cm}
Galaxy formation increases the entropy of intracluster gas, producing
a relationship with temperature that is flatter than the self-similar
scaling ($S \propto T$). We plot this relation in Fig.~\ref{fig:st},
again using an X-ray emission-weighted temperature for each cluster.
Two radii are considered ($0.1r_{200}$ and $r_{500}$) 
and only clusters with $kT_{\rm X}>1$keV are studied. Again,
the simulated clusters are in reasonably good agreement with the
observations \citep{Ponman03}, containing an excess of entropy that is
larger in smaller systems. 
For $0.1r_{200}$
\begin{equation}
\log(S/\kevcm2) = (2.14 \pm 0.008) + (0.46 \pm 0.03) \, \log(kT_{\rm X}/\keV)
\label{eqn:st_0.1r200}
\end{equation}
and for $r_{500}$
\begin{equation}
\log(S/\kevcm2) = (2.84 \pm 0.003) + (0.63 \pm 0.01) \, \log(kT_{\rm X}/\keV).
\label{eqn:st_r500}
\end{equation}

\subsection{Luminosity--temperature relation}
\label{subsec:lt}

\begin{figure}
\centering
\centerline{\epsfig{file=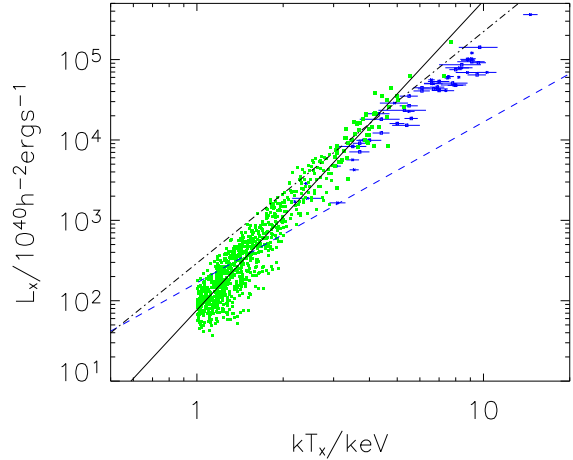,width=3in}}
\caption{Bolometric X-ray luminosity versus emission-weighted
temperature in the 1.5-8.0 keV band. Symbols with error bars
are data from Markevitch (1998) and Arnaud and Evrard (1999).
The dashed line is the best-fit relation for a fixed slope of 2.
The solid line is the best-fit relation for $kT_{\rm X}>1$keV
and the dot-dashed line for $kT_{\rm X}>3$keV.}
\label{fig:lt}
\end{figure}

\vspace{-0.5cm} Finally, we show bolometric X-ray luminosity versus
X-ray emission-weighted temperature in Fig.~\ref{fig:lt}. Again, only
clusters with $kT_{\rm X}>1$keV are considered. Symbols with errors
are observational data from \citet{Markevitch98} and
\citet{Arnaud99}. To remain approximately consistent with this data,
emission from within $50\hkpc$ of our simulated cluster centres is
omitted.

The dashed line is a best-fit relation for a fixed slope equal to 2
(self-similar), clearly a poor fit to the observations. When the slope
is allowed to vary, the best-fit relation (solid line) is
\begin{equation}
\log(L_{\rm X}/L_{40}) = (1.89 \pm 0.01) + (3.84 \pm 0.05) \, \log(kT_{\rm X}/\keV),
\label{eqn:lxtx_bf}
\end{equation}
where $L_{40}=10^{40}\Lxunit$, considerably steeper than the
self-similar case. In fact, the simulated relation is not adequately
described by a power law since the local gradient becomes
progressively flatter with increasing temperature. Fitting clusters
with $kT_{\rm X}>3$keV (dot-dashed line) yields
\begin{equation}
\log(L_{\rm X}/L_{40}) = (2.47 \pm 0.03) + (2.88 \pm 0.05) \, \log(kT_{\rm X}/\keV),
\label{eqn:lxtx_bf3}
\end{equation}
in reasonable agreement with the observations although the
normalisation is a bit too high. Better agreement was reached by
\citet{Kay04}, who used a slightly smaller gas fraction (0.15 rather
than 0.162 used here, which is a closer match to the \WMAP value). It
is likely however that fine tuning of the feedback model parameters
would improve the agreement between the simulated and observed
relations.





\end{document}